\documentclass[useAMS,usenatbib]{mn2e}

\usepackage{graphicx}
\usepackage{epstopdf}
\usepackage{epsfig}
\usepackage{enumerate}
\usepackage{amsmath,amssymb}

\title[Using the cosmic bulk flow to constrain $f(R)$ gravity]{Using measurements of the cosmic bulk flow to constrain $f(R)$ Gravity}
\author[Seiler \& Parkinson]{Jacob Seiler$^{1,2}$\thanks{E-mail:
jacob.seiler@uq.net.au}, David Parkinson$^{1}$\\
\\
$^{1}$School of Mathematics and Physics, University of Queensland, Brisbane, QLD 4072, Australia \\
$^{2}$ARC Centre of Excellence for All-Sky Astrophysics (CAASTRO)}
\begin{document}


\pagerange{\pageref{firstpage}--\pageref{lastpage}} \pubyear{2002}

\maketitle

\label{firstpage}

\begin{abstract}

As an alternative explanation for the cosmic acceleration, $f(R)$ theories of gravity can predict an almost identical expansion history to standard $\Lambda$CDM, yet make very different predictions for the growth of cosmological structures. Measurements of the cosmic bulk flow provides a method for determining the strength of gravity over the history of structure formation. We use the modified gravity N-body code ECOSMOG to simulate dark matter particles and make predictions for the bulk flow magnitude in both $\Lambda$CDM and $f(R)$ gravity.  With the peculiar velocities output by ECOSMOG we determine the bulk flow at depths ranging from $20h^{-1}$Mpc to $50h^{-1}$Mpc, following the redshift and sky distribution of the 2MASS Tully-Fisher survey (2MTF).  At each depth, we find that the $\Lambda$CDM and $f_{R0} = 10^{-5}$ simulations produce bulk flow measurements that are consistent with $\Lambda$CDM predictions and the 2MTF survey at a $1\sigma$ level. We also find that adopting an $f(R)$ strength of $f_{R0} = 10^{-3}$ predict a much larger value for the bulk flow, which disagree with $\Lambda$CDM predictions at all depths considered.  We conclude that $f_{R0}$ must be constrained to a level no greater than $10^{-4}$ to agree with bulk flow measurements.
\end{abstract}


\section{Introduction}

One of the most inescapable facts in recent cosmology is that the Universe is undergoing a period of accelerated expansion.  The effect of this acceleration was observed through measurements of supernovae \citep{b1,b2}, confirming previous indications from large-scale structure and galaxy surveys \citep{Efstathiou,OstrikerSteinhardt,KraussTurner,YoshiiPetersen}.
The source of this late-time acceleration has been named `Dark Energy' which exerts a negative pressure to combat the attractive force of gravity.


Currently the simplest candidate for dark energy is the cosmological constant $\Lambda$.  However theoretical calculations yield a value of $\Lambda$ at least 120 orders of magnitude larger than observations \citep{b4,b5}. As a result, cosmological models that do not include an explicit cosmological constant form an appealing alternative.  These alternatives are usually categorized depending upon which side of the Einstein equations they alter.  The first category adds to or alters the energy-momentum tensor $T_{\mu\nu}$ to yield a negative pressure (dark fluid models), while the second category alters the Einstein tensor $G_{\mu\nu}$ to generate the acceleration (modified gravity models).  Throughout this paper we focus on a specific modified theory, $f\left(R\right)$ gravity.

$f\left(R\right)$ gravity changes the gravitational theory by modifying the action, from the standard Einstein-Hilbert action, to be some new function of the Ricci scalar $R$ \citep{Nojiri,Carroll}. Given the freedom to choose the function $f\left(R\right)$,  the expansion histories of both $\Lambda$CDM and $f\left(R\right)$ models can be very similar, or even identical \citep{Song}. Therefore we must consider alternate methods to observationally differentiate between the models.  One such approach is to study the peculiar velocity of galaxies which result from the gravitational interaction between a galaxy and the surrounding matter, causing the galaxy redshift to deviate from Hubble's Law.  In essence, the peculiar velocity of a galaxy is an integrated history of its gravitational interactions, and thus provides a tool to differentiate between $\Lambda$CDM and $f\left(R\right)$ models.

Measuring peculiar velocities offers an observational difficulty as such measurements must be performed using redshift independent distance indicators such as type Ia Supernovae \citep{b12}, the Tully-Fisher relation \citep{b13} and the Fundamental Plane relation \citep{b14}. A common parameter that many peculiar velocity surveys quote is the net dipole, or the `bulk flow', of the peculiar velocity field.  There has been much debate over whether the measured bulk flows are consistent with the $\Lambda$CDM model.  \cite{b15} analyzed 2,018 galaxies from the 2MASS Tully-Fisher survey (2MTF) utilizing both $\chi^2$ and minimum variance methods, finding a bulk flow that is consistent with the $\Lambda$CDM model to a $1\sigma$ level.  Conversely, \cite{b16} utilized a catalogue of 4,481 peculiar velocity measurements with a characteristic depth of $33h^{-1}$Mpc and claim that the resulting bulk flow is inconsistent with the $\Lambda$CDM model at a $>$98\% confidence level.  
\newline \indent A possible solution to these anomolaous bulk flow measurements is to adopt a modified theory of gravity. To this end, N-body simulations can be employed to evolve particles under both $\Lambda$CDM and modified gravity models.  The results of these simulations can then be compared to surveys such as 2MTF.  An added benefit of utilizing N-body simulations to measure bulk flow is the lack of underlying systematic biases that most surveys are subject to.  This is especially important as \cite{b17} has shown that unaccounted systematic uncertainty could explain the discrepancies between surveys agreeing/disagreeing with the $\Lambda$CDM model. 
\newline \indent In this paper we utilize N-body simulations to measure bulk flow in both $\Lambda$CDM and $f\left(R\right)$ regimes.  In Section 2 we outline $f\left(R\right)$ gravity and show how we quantify the deviation from the $\Lambda$CDM model.  In Section 3 we give an overview of the simulations we use, how the output is utilized to calculate bulk flow and a brief outline of the 2MTF survey.  In Section 4 we present the results of the simulations and compare them to the 2MTF survey.  We conclude in Section 5.  
\newline \indent Throughout the paper we adopt a standard cosmology of $\Omega_m = 0.30$, $\Omega_\Lambda = 0.70$ and $H_0 = 100h$ km s$^{-1}$Mpc$^{-1}$.  Whilst our results are $h$ independent, we use a value of $h = 0.70$ in our simulations. 

\section{Modified Gravity \label{sec:MG}}

Dynamics in a General Relativistic regime are governed by the Einstein-Hilbert action given by
\begin{equation}
S = \frac{1}{16\pi G}\int\limits_{}^{} d^4x \sqrt{-g}\left(R + f\left(R\right)\right) + S_m\left(g_{\mu\nu}, \Psi_m\right),
\label{eq:ActionHu}
\end{equation}
\noindent where $G$ is the universal gravitational constant, $g$ is the determinant of the metric $g_{\mu\nu}$, $f(R)$ is some general function of the Ricci scalar R, $S_m$ is the action of some matter fields $\Psi_m$ and we have used units where $c = 1$. 


By varying the action with respect to the metric we obtain
\begin{equation}
G_{\mu\nu} +f_RR_{\mu\nu} -\nabla_\mu \nabla_\nu f_R - \left(\frac{f\left(R\right)}{2} - \Box f_R\right)g_{\mu\nu} = 8\pi GT_{\mu\nu},
\label{eq:Varying}
\end{equation}
\noindent where the field $f_R = \frac{\partial f(R)}{\partial R}$, $\Box = \partial^\mu\partial_\mu$ is the D'Alembert operator and $T_{\mu\nu}$ is the energy-momentum tensor.  If we select $f(R)$ according to $\Lambda$CDM, $f(R) = -2\Lambda$, we see that the derivatives in equation \ref{eq:Varying} vanish recovering the Einstein field equation.  In the $f\left(R\right)$ regime, the N-body code that we use (see Section 3) employs the expression
\begin{equation}
f\left(R\right) = - m^2\frac{c_1\left(-R/m^2\right)^n}{c_2\left(-R/m^2\right)^n + 1},
\label{eq:fr}
\end{equation}
\noindent where $n > 0$, $c_1$ and $c_2$ are model parameters and $m^2 = \Omega_mH_0^2$ is the characteristic length scale, with $\Omega_m$ being the present fractional matter density \citep{b20}.  

By definition of this modified theory of gravity, there is no true cosmological constant.   However, at curvatures larger than $m^2$, $f\left(R\right)$ may be expanded as
\begin{equation}
\lim\limits_{m^2/R \rightarrow 0} f\left(R\right) \approx -\frac{c_1}{c_2}m^2 + \frac{c_1}{c_2^2}m^2\left(\frac{m^2}{R}\right)^n.
\end{equation}
The limiting case of $\frac{c_1}{c_2^2} \rightarrow 0$, at fixed $\frac{c_1}{c_2}$, is a cosmological constant hence our model requires that as $\frac{c_1}{c_2^2} \rightarrow 0$, we approach $\Lambda$CDM gravity.  Furthermore, by taking the trace of equation \ref{eq:Varying}, one obtains a field equation for $f_R$
\begin{equation}
3\Box f_R - R + f_RR - 2f\left(R\right) = -8\pi G\rho,
\end{equation}
\noindent where $\rho$ is the density of the Universe.  As a result, the impact of $f\left(R\right)$ gravity can be viewed in terms of the field $f_R$ \citep{b20}. Utilizing these two facts, we characterize the deviation from the $\Lambda$CDM model by the value of $f_R$ at the present epoch given by
\begin{equation}
f_{R0} \approx -n\frac{c_1}{c_2^2}\left(\frac{12}{\Omega_m} -9\right)^{-n-1},
\end{equation}
\noindent where larger $n$ mimics $\Lambda$CDM until later times. Throughout our work we exclusively use $n = 1$.

\section{Simulation Outline and Methods \label{sec:Sims}}

\subsection{Simulation Outline}

For all our simulations we use $512^3$ dark matter particles placed inside a box with side length $500h^{-1}$Mpc.  The initial conditions were generated using 2LPTic which uses second order Lagrangian perturbation theory thereby offering increased accuracy compared to the Zel'dovich approach \citep{b21}.  To specify how the 2LPTic grid is distributed, we use the transfer function and power spectrum at $z = 0$ generated by CAMB \citep{b22} to calculate the power spectrum at the starting redshift $z = 49$.  The $\sigma_8$ value used is $0.7911$ assuming a $15\%$ baryon fraction.

In this work we use the N-body simulator ECOSMOG\footnote{We obtained a copy of the ECOSMOG code from the authors, and used it with permission.} which is specifically designed to simulate the universe under $f\left(R\right)$ gravity \citep{b23}.  ECOSMOG is based upon the RAMSES code, and uses an adaptive mesh refinement (AMR), which allows direct control of the trade off between accuracy and speed \citep{b24}.  We have a minimum force resolution of $0.97h^{-1}$Mpc (for the course grid), increasing to a maximum of $15h^{-1}$kpc (with six levels of refinement), which is sufficient for our work where our minimum sphere radius is $20h^{-1}$Mpc. The code works by locally solving the perturbation equations for the gravitational potential $\Phi$ and $f_R$ field
\begin{align}
\nabla^2\Phi & = \frac{16\pi G}{3}a^2\delta \rho_M + \frac{a^2}{6}\delta R\left(f_R\right), \label{eq:1} \\
\nabla^2\delta f_R & = -\frac{a^2}{3}\left[\delta R\left(f_R\right) + 8\pi G \delta \rho_M\right], \label{eq:2}
\end{align}
where $\delta f_R = f_R\left(R\right) - f_R\left(\bar{R}\right)$, $\delta R = R - \bar{R}$, $\delta \rho_M = \rho_M - \bar{\rho}_M$ and the overbars denote the background values \citep{b23}.  In underdense regions, the $\delta R\left(f_R\right)$ in equation \ref{eq:1} vanishes causing the two equations to decouple and resulting in gravity simply being enhanced by a factor of $4/3$.  However in overdense regions, $\delta f_R$ becomes negligible recovering the Poisson equation for general relativity, $\delta R\left(f_R\right) = -8\pi G\delta \rho_M$.

\subsection{2MTF Survey}

The Two Micron All-Sky Survey (2MASS; \citealp{b25}) Tully-Fisher Survey (2MTF; \citealp{b27}) utilizes photometry data from 2MASS in conjunction with rotation and HI widths to calculate the Tully-Fisher (TF) distance from the redshifts in the 2MASS Redshift Survey (2MRS; \citealp{b26}). In essence, 2MTF worked to calculate a universal calibration for the TF relation when utilizing the 2MASS photometry data in the $J$, $H$ and $K$ bands. The difficulty in deriving a global TF calibration is that thought must be given to survey specific biases as explained in depth by \cite{b27}. \cite{b15} then calculate peculiar velocities of $2018$ galaxies by comparing the magnitude predicted by the observed redshift of the galaxy and the TF predicted redshift.  

With the peculiar velocities calculated, 2MTF determines the bulk flow at various depths by applying a $\chi^2$ minimization using weights that account for the measurement error in distance ratio and the redshift and number density distributions of the galaxies. \cite{b15} also calculates the bulk flow via other methods which we do not use for comparison in this paper for simplicity's sake.

We use the 2MTF survey as a comparison as it covers both small and large scales allowing us to probe how $f\left(R\right)$ bulk flows compare at a variety of depths.  It is possible to extend our work to even larger scales utilizing other work such as \cite{b28} who use data from the 6dF Galaxy Survey to calculate bulk flows of depths from $50$ to $70$ $h^{-1}$Mpc. This further comparison was not performed in this work as the data was not available at the time this research was undertaken.

\subsection{Mock Surveys}

We run the simulation until $z = 0$ (in a co-moving frame) at which point the simulator outputs the position and velocity of each dark matter particle.  As the equations of motion are solved in a co-moving frame, the output include only the peculiar velocity of the particle.

\begin{figure*}
	\includegraphics[scale = 0.13]{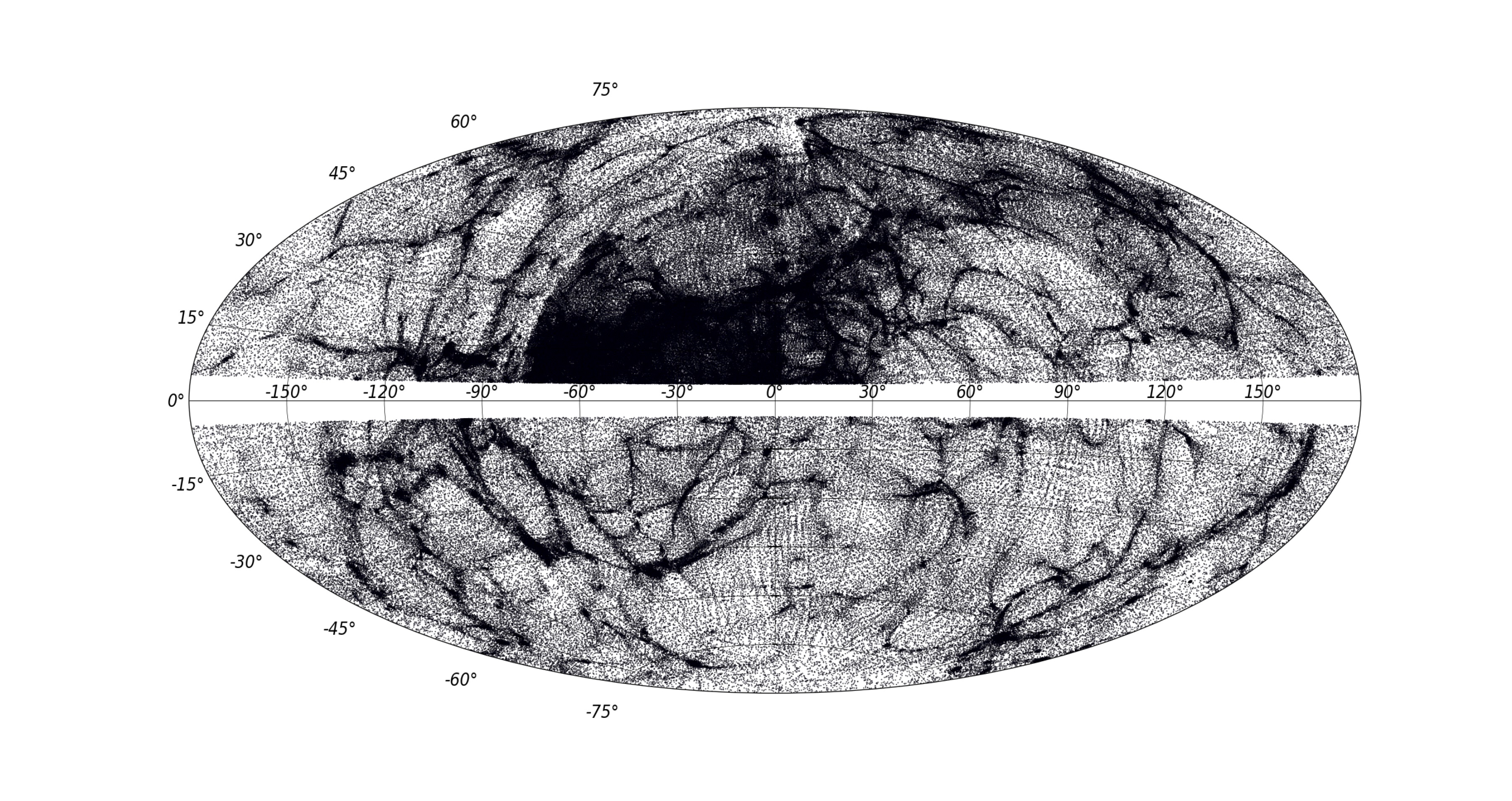}
	\caption{Aitoff projection in galactic coordinates of accepted particle positions for a single mock survey.  We mimic 2MTF in that particles with latitude $\left|b\right| < 5^\circ$ are excluded \citep{b15}.}
	\label{fig:Angles}
\end{figure*}

\begin{figure}
	\includegraphics[scale = 0.45]{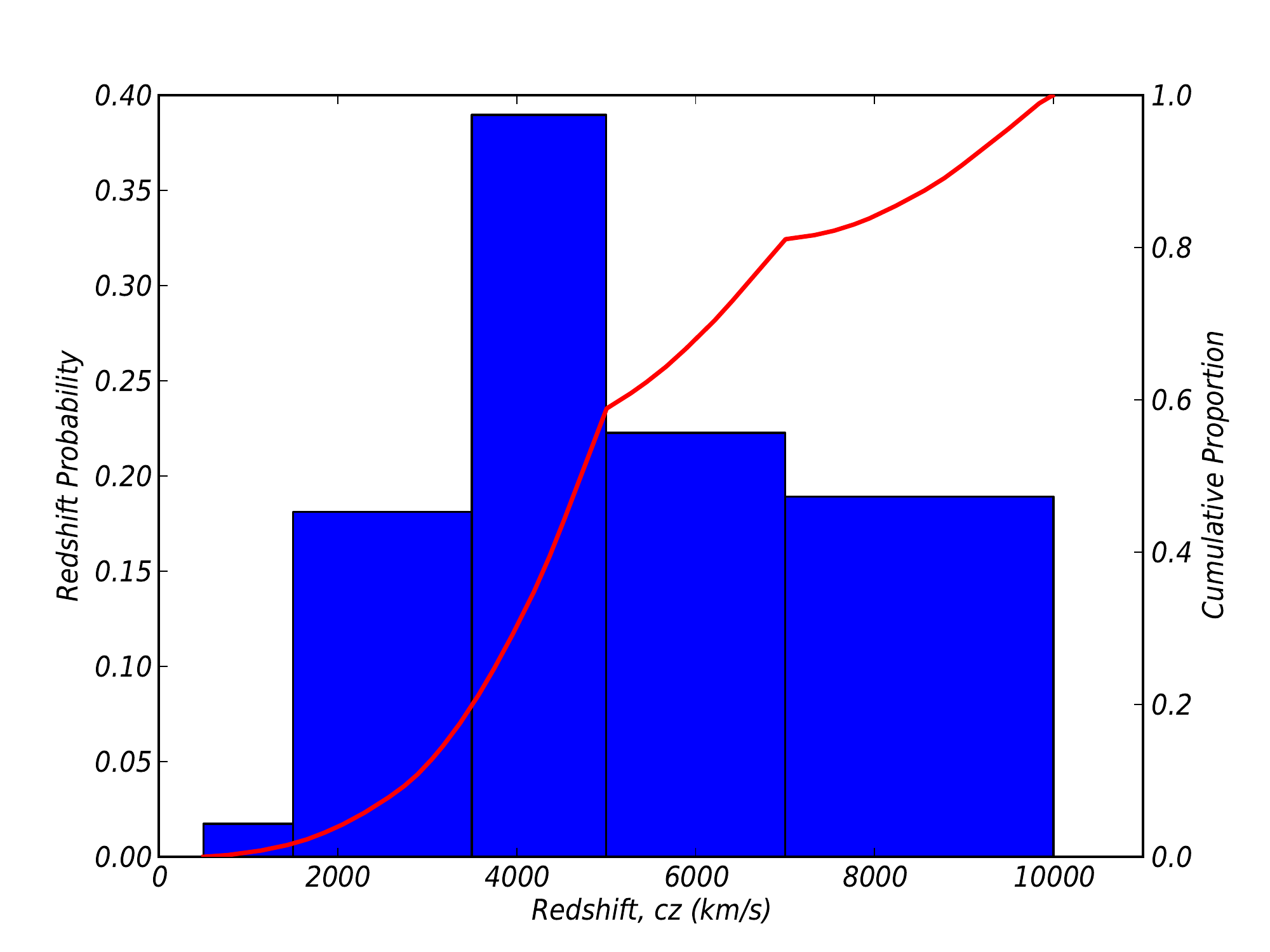}
	\caption{Mean redshift distribution (histogram left axis, cumulative proportion solid line right axis) over $50$ mocks.  This distribution was selected to closely follow the 2MTF survey \citep{b15}.}
	\label{fig:Histogram}
\end{figure}

The bulk flow in a spherical region of radius $r$ is given by 
\begin{equation}
\textbf{B}\left(r\right) = \frac{3}{4\pi r^3}\int\limits_{x<r}^{}\textbf{v}\left(\textbf{x}\right)d^3x,
\label{eq:BulkDef}
\end{equation}
where $\textbf{v}\left(\textbf{x}\right)$ is the peculiar velocity field.  Due to the nature of this equation, the peculiar velocity must be sampled uniformly over the volume.  However as we wish to compare our results to 2MTF in which the uniformity requirement is not met, we calculate the bulk flow via the following process:  

\begin{enumerate}
	\item {Select a random point inside the simulation box to be the center of the mock.  This point is chosen such that any particle in the mock survey lies within the box.  For our work, we use the 2MTF distance bound of $100h^{-1}$Mpc.  We further select the random point such that spheres of radius $50h^{-1}$Mpc will not overlap each other ensuring that each mock survey is independent.}
	\item {For each particle, if it is within the distance and latitude bounds, bin the particle into the corresponding redshift bin.  This redshift is the observed redshift and is calculated using equation \ref{eq:Redshift}, where $z_{\text{pec}}$ is the peculiar redshift determined by projecting the peculiar velocity along the line of sight and $z_{\text{rec}}$ is the recession redshift dictated by the co-moving distance between the particle and center of the mock. Once again we follow 2MTF which only surveyed galaxies with latitude $\left|b\right| > 5^\circ$ (Figure \ref{fig:Angles}).}
	\begin{equation}
	1+ z_{\text{obs}} = (1 + z_{\text{pec}})(1 + z_{\text{rec}})
	\label{eq:Redshift}
	\end{equation}
	\item {Normalize the redshift histogram and create an array of accepted particles such that the resulting distribution will follow that of 2MTF (Figure \ref{fig:Histogram}).  The accepted particles are chosen randomly from each redshift bin.  Note that the similarity between the redshift distribution of the mocks and 2MTF is not overly important and will not affect the results in a significant manner; as such, we only roughly follow the 2MTF distribution.}
	\item {For each sphere radius, if an accepted particle lies within the radius, add its peculiar velocity component to the total.  Once all accepted particles have been checked, the bulk flow is given by equation \ref{eq:Bulk} where $B_x,B_y,B_z$ is the net peculiar velocity in each direction and $N$ is the number of particles inside the sphere of radius $r$.}
	\begin{equation}
	\left|\textbf{B}\left(r\right)\right| = \frac{\sqrt{B_x^2 + B_y^2 + B_z^2}}{N}  
	\label{eq:Bulk}
	\end{equation}
\end{enumerate}


\section{Results \label{sec:Results}}

\begin{figure}
	\includegraphics[scale = 0.45]{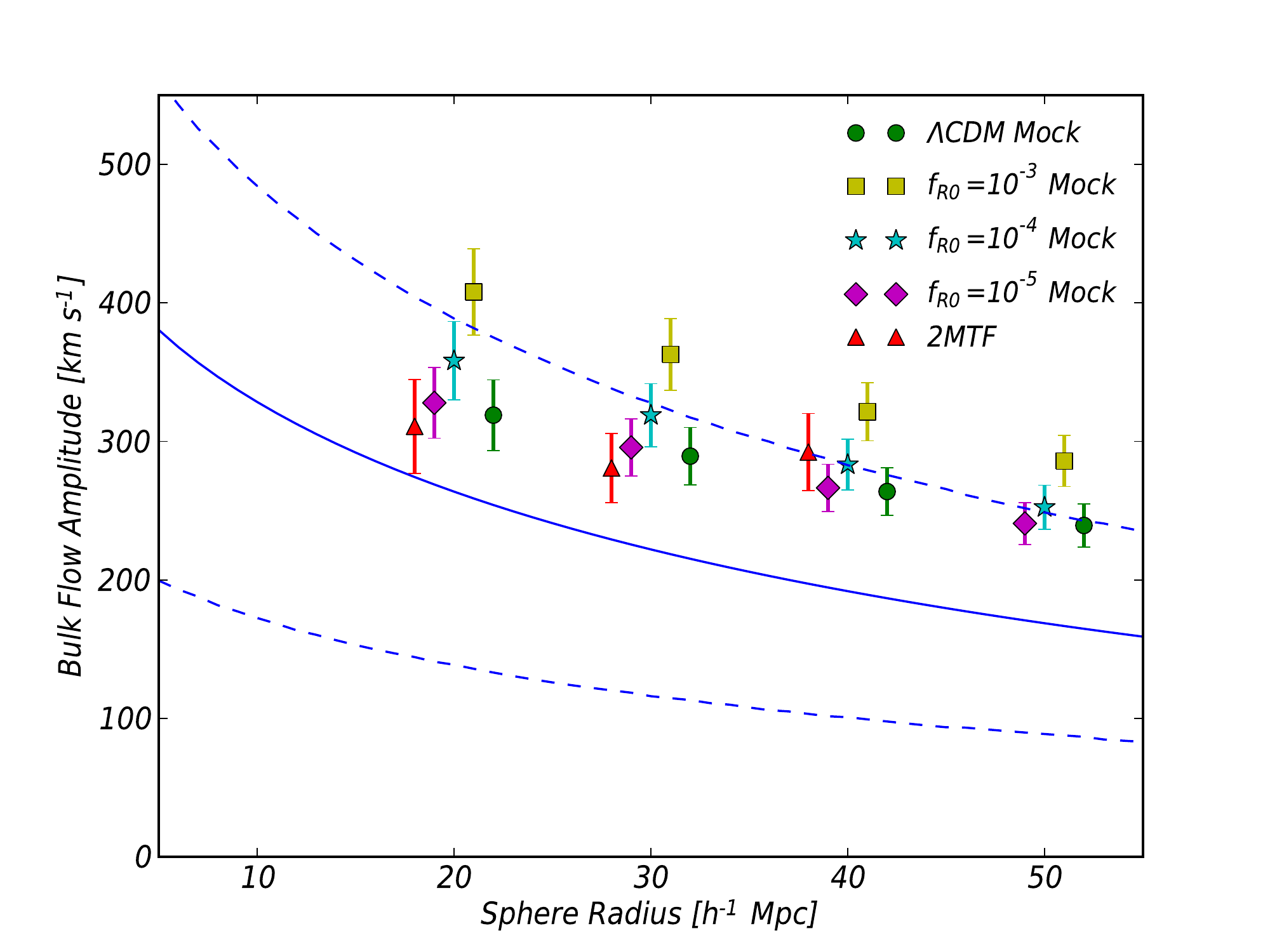}
	\caption{Average bulk flow magnitude over $50$ mock surveys in $\Lambda$CDM gravity (circles), $f_{R0} = 10^{-3}$ (squares), $f_{R0} = 10^{-4}$ (stars) and $f_{R0} = 10^{-5}$ (diamonds).  The error bars denote the $1\sigma$ scatter of the mocks.  The solid line indicates the $\Lambda$CDM prediction with a $1\sigma$ uncertainty shown as the dashed lines.  For comparison we list the 2MTF 3-band $\chi^2$ minimization result as triangles \citep{b15}. Note that the sphere radius is the same for each data set (from $20h^{-1}$Mpc to $50h^{-1}$Mpc in intervals of $10h^{-1}$Mpc) but have been shifted on this plot for better visibility.}
	\label{fig:Results}
\end{figure}

Figure \ref{fig:Results} and Table \ref{tab:Results} shows the mean bulk flow amplitude over $50$ mock surveys in both $\Lambda$CDM and $f\left(R\right)$ regimes at depths ranging from $20h^{-1}$Mpc to $50h^{-1}$Mpc.  As a point of reference, we also show the 2MTF results for their 3 band, $\chi^2$ minimization, at depths of $20h^{-1}$Mpc, $30h^{-1}$Mpc and $40h^{-1}$Mpc \citep{b15}.  $50$ mocks provides adequate convergence for the covariance matrix for the uncertainty in the bulk flow amplitude (Appendix \ref{sec:Error}) 

Following the procedure outlined in \cite{b15}, the $\Lambda$CDM bulk flow variance is given by
\begin{equation}
v_{rms}^2 = \frac{H_0^2f^2}{2\pi^2}\int\limits_{}^{}W^2\left(kR\right)P\left(k\right)dk,
\label{eq:rms}
\end{equation}
where $H_0$ is the Hubble constant,  $f = \Omega_m^{0.55}$ is the linear growth rate, $W\left(kR\right) = \exp\left({-k^2R^2}/2\right)$ is the Gaussian window function, $k$ is the wavenumber and $P\left(k\right)$ is the matter power spectrum.

The probability density function for a bulk flow amplitude B is given by 
\begin{equation}
p\left(B\right)dB = \sqrt{\frac{2}{\pi}}\left(\frac{3}{v_{rms}^2}\right)^{3/2}B^2\exp{\left(-\frac{3B^2}{2v_{rms}^2}\right)}dB,
\label{eq:Distribution}
\end{equation}
where the distribution has been normalized by setting $dp(B)/dB = 0$ \citep{b25}. Due to the nature of bulk flow, this distribution is Maxwellian hence the peak will occur at $\sqrt{2/3}v_{rms}$.  We adopt this peak as the theoretical bulk flow measurement in the $\Lambda$CDM regime with $1\sigma$ error bars given by integrating equation \ref{eq:Distribution} around the peak.

From Table \ref{tab:Results} we observe that there is little difference between $\Lambda$CDM and $f_{R0} = 10^{-5}$ bulk flow amplitudes.  This is not too surprising as the $f\left(R\right)$ modification is negligible recovering standard $\Lambda$CDM.  There is a noticeable increase in the bulk flow amplitude when the $f\left(R\right)$ strength is increased to $f_{R0} = 10^{-3}$ or $10^{-4}$.  This is the result of the Poisson equation being enhanced by the presence of a non-negligible $f_R$.

We find that our $f_{R0} = 10^{-5}$ and $\Lambda$CDM results agree comfortably with the 2MTF survey at all scales.  Whilst the 2MTF result at $40h^{-1}$Mpc agrees with all of our mocks, it disagrees with the expected trend of decreasing amplitude as sphere radius increases suggesting that we should be hesitant to take it to be as an accurate data point in comparison to our mocks.  The $f_{R0} = 10^{-5}$ and $\Lambda$CDM mocks are in agreement with the theoretical predictions for all scales with the $f_{R0} = 10^{-4}$ results agreeing only at smaller scales.  In essence, we are constraining $f_{R0}$ to be below $10^{-3}$ and between $10^{-4}$ and $10^{-5}$ at a one sigma level.   Such a constraint matches previous work where $f_{R0}$ was constrained to $<6 \times 10^{-5}$, $<1.3 \times 10^{-3}$, $<3.5 \times 10^{-3}$ and $<1.3 \times 10^{-4}$ levels \citep{b30, b28,b29,b27}.  Furthermore if we wished to explain the anomolaous result of \cite{b16} who found a bulk flow amplitude of $407 \pm 81$ km s$^{-1}$ (error to $3\sigma$) on a scale of $50h^{-1}$Mpc, we would need to adopt $f_{R0} > 10^{-3}$ which disagrees with both 2MTF and the previously cited work.


\begin{table*}
	\begin{center}
		\begin{tabular}{|c|c|c|c|c|c|}
			\hline
			\textbf{Sphere Radius ($h^{-1}$Mpc)} & \textbf{$\Lambda$CDM (km/s)} & \textbf{$f_{R0} = 10^{-5}$ (km/s)} &  \textbf{$f_{R0} = 10^{-4}$ (km/s)} & \textbf{$f_{R0} = 10^{-3}$ (km/s)} & \textbf{2MTF} \\ \hline \hline
			20 & 319.0 $\pm$ 25.5 & 327.8 $\pm$ 25.6 & 358.2 $\pm$ 28.4 & 408.0 $\pm$ 31.2 &  310.9 $\pm$ 33.9 \\
			30 & 289.3 $\pm$ 20.7 & 295.6 $\pm$ 20.6 & 319.0 $\pm$ 22.7 & 362.8 $\pm$ 25.9 &  280.8 $\pm$ 25.0 \\ 
			40 & 263.8 $\pm$ 17.3 & 266.4 $\pm$ 17.1 & 283.2 $\pm$ 18.3 & 321.5 $\pm$ 21.0 &  292.3 $\pm$ 27.8 \\
			50 & 239.2 $\pm$ 15.5 & 240.7 $\pm$ 15.1 & 252.5 $\pm$ 15.9 & 285.8 $\pm$ 18.3 & - \\
			\hline
		\end{tabular}
		\caption{Average flow magnitude for various gravitational models over $50$ mock surveys and 2MTF survey \citep{b15}.  Uncertainty is given to a $1\sigma$ level and for our mocks is determined by calculating the scatter in the bulk flow amplitude.}
		\label{tab:Results}
	\end{center}
\end{table*}

\section{Conclusion \label{sec:Conclusion}}

Modified gravity theories provide an appealing alternative to the $\Lambda$CDM model by providing a model that does not include an explicit cosmological constant.  One such theory, $f\left(R\right)$ gravity, involves changing the Einstein-Hilbert action by altering the functional dependence upon the Ricci scalar. We consider one particular $f(R)$ model, the Hu \& Sawicki model \citep{b20}, and parameterise the degree of deviation from the predictions of standard $\Lambda$CDM through the gradient of the function today $f_{R0}=\frac{\partial f(R)}{\partial R}|_{t=t_0}$.

The bulk flow is the net dipole moment of the cosmological peculiar velocity field, which is the result of gravitational influence on the motions of particles on large scales.  As the bulk flow is sensitive to the gravitational theory considered, this results in a measurable difference in the predicted value of the bulk flow between $\Lambda$CDM and $f\left(R\right)$ gravity.  

We utilized N-body simulations to create a set of mock surveys under both $\Lambda$CDM and $f\left(R\right)$ gravity.  These mocks were analyzed assuming the redshift and sky distribution of 2MTF, a previous survey that studied the bulk flow in the local Universe \citep{b15}.  We found that the simulations under $\Lambda$CDM and $f_{R0} = 10^{-5}$ gravity produced bulk flows that were consistent with both 2MTF and $\Lambda$CDM predictions.  Choosing $f_{R0} = 10^{-4}$ gave bulk flows that agreed with the $\Lambda$CDM predictions on small scales with weak agreement on large scales.  Finally, setting $f_{R0} = 10^{-3}$ resulted in bulk flows that did not agree with $\Lambda$CDM predictions to a $1\sigma$ level at all scales considered.  From these results we conclude that in order to obtain bulk flow measurements that match previous work \citep{b30, b28,b29,b27} in addition to theoretical predictions, the upper limit on $f_{R0}$ lies somewhere in the range $10^{-4}$ and $10^{-5}$.

We finally note that, given the agreement between the $\Lambda$CDM and $f_{R0} = 10^{-5}$ predictions, it seems unlikely that bulk flow measurements can be of any further use in constraining the parameters in an $f(R)$ theory. We have already reached the theoretical limit in which the bulk flow amplitude will provide useful information. Instead to make further progress in this area, it is more advantageous to use the full velocity power spectrum \citep[e.g.][]{Johnson}.


\section*{Acknowledgements}

We thank Tamara Davis, Baojiu Li, Chris Springob and Chris Power for helpful comments and suggestions while this work was under preparation. Parts of this research were conducted by the Australian Research Council Centre of Excellence for All-sky Astrophysics (CAASTRO), through project number CE110001020. Parts of the computational analyses were supported by the Flagship Allocation Scheme of the NCI National Facility at the ANU. JS acknowledges the hospitality of the International Centre for Radio Astronomy at the University of Western Australia for part of this research. DP is supported by an  Australian Research Council Future Fellowship [grant number FT130101086].

\section{Appendix}

\subsection{Line of Sight Vs. $3$D Velocity}

Throughout most of the literature, authors simply state that the bulk flow in one dimension is a factor of $1/\sqrt{3}$ smaller than the full three dimensional amplitude \citep{b26, b16}.  However the reasoning behind this factor is not fully explored.  In this appendix we wish to briefly give an overview of the logic behind the $1/\sqrt{3}$ factor difference between line of sight and three dimensional bulk flow amplitude.

Consider particles with velocity $\textbf{v}_{3D} = \left(v_x, v_y, v_z\right)$.  We choose to position our coordinate system such that one of the axes lies precisely along the line of sight yielding a velocity $\textbf{v}_{LoS} = v_x$.  Then the ratio of the three dimensional and the line of sight bulk flows in such a situation is given by

\begin{align}
\frac{\textbf{B}_{3D}\left(r\right)}{\textbf{B}_{LoS}\left(r\right)} & = \sqrt{\frac{v_x^2 + v_y^2 + v_z^2}{v_x^2}}, \\
& = \sqrt{1 + \frac{v_y^2 + v_z^2}{v_x^2}} \label{eq:Ratio}.
\end{align}

\begin{figure}
	\includegraphics[scale = 0.45]{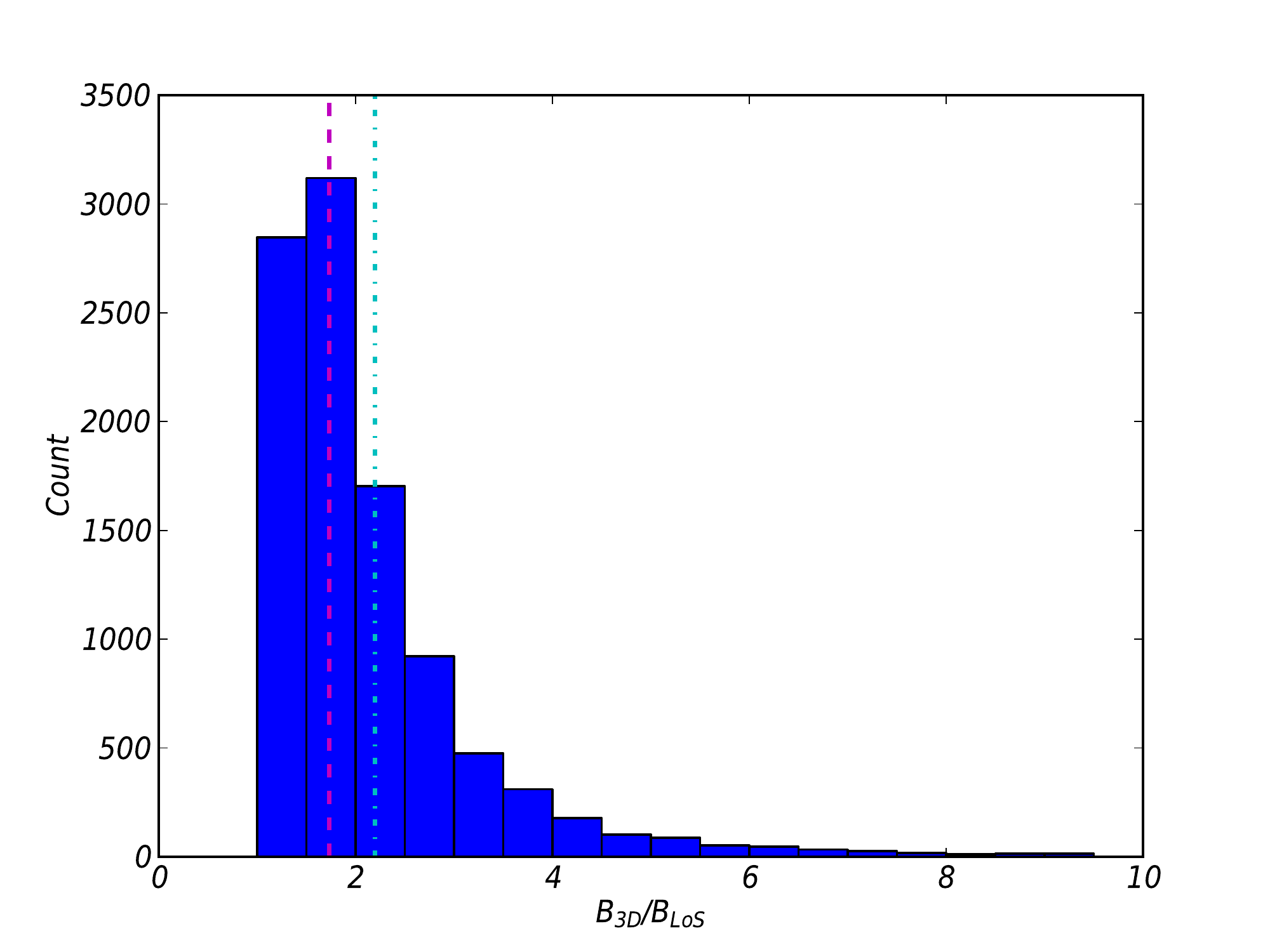}
	\caption{Distribution for the ratio between three dimensional and line of sight bulk flows.  The dashed line is at $\sqrt{3}$ which matches literature and the dash-dot line depicts the arithmetic mean of the distribution with a value of $2.19$.}
	\label{fig:LoS}
\end{figure}

These velocities are drawn from a Maxwellian distribution with the same variance.  In Figure \ref{fig:LoS} we plot a histogram of $10,000$ ratios following equation \ref{eq:Ratio}.  We see that the peak of this distribution is centered on $\sqrt{3}$ despite the arithmetic mean occurring at a higher value of $2.19$.  This is due to $v_x \ll v_y, v_z$ case skewing the ratio towards extremely high values.  Thus the 1-dimensional amplitude of the bulk flow is not the arithmetic mean of the individual particle contributions (as might be assumed), but rather the peak of the distribution.

\subsection{Uncertainty Convergence \label{sec:Error}}

\begin{figure}
	\includegraphics[scale = 0.4]{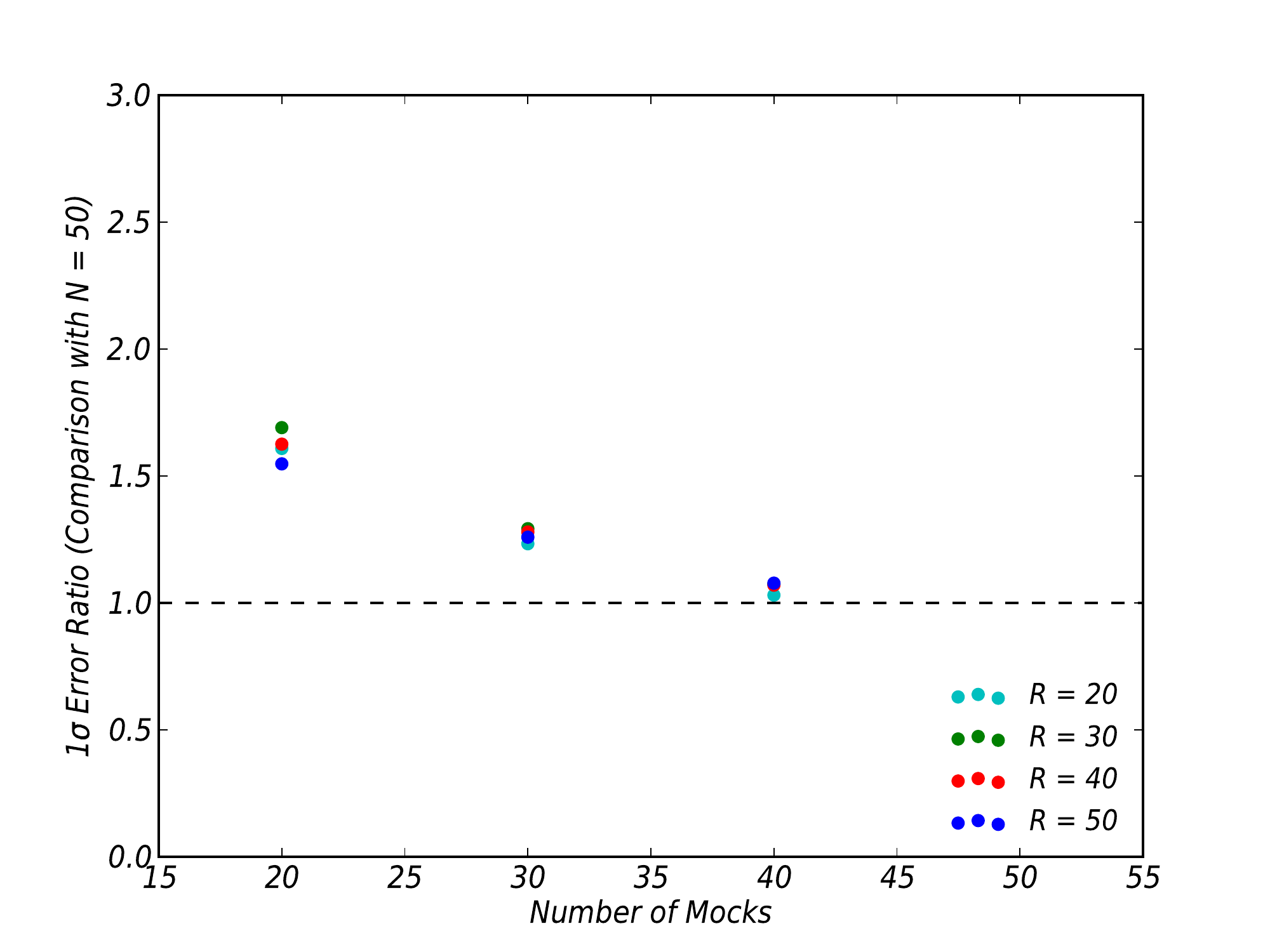}
	\caption{Ratio of the $1\sigma$ uncertainty in the bulk flow amplitude for the $\Lambda$CDM simulation (with reference to $N = 50$ mocks) for different mock numbers (left) and different radii (right, in units of $h^{-1}$Mpc).}
	\label{fig:Errors}
\end{figure}

In this appendix we show that the number of mocks ($N = 50$) provides adequate convergence for the uncertainty in the bulk flow amplitudes.  We do this by varying the number of mocks from $10$ to $50$ and plotting the ratio of the $1\sigma$ uncertainty (with reference to $N = 50$ mocks) in Figure \ref{fig:Errors}.  We see that as the number of mocks approaches $50$ we approach convergence, and this is true for every sphere radius. 

\end{document}